\documentstyle[aps,pra,epsf,preprint,tighten]{revtex}
\input epsf.sty
\begin{document}
\widetext
\title{Non-Fermi liquid behavior in U and Ce intermetallics}
\author{A.~H.~Castro Neto}
\bigskip
\address
{Department of Physics,
University of California,
Riverside, CA 92521}
 
\maketitle

\begin{center}
\tt Proceedings of the \\
VI International Summer School ``Nicol\'as Cabrera''\\
Miraflores de la Sierra, Madrid \\
September 14-18, 1999.
\end{center}
 
\begin{abstract}
In this paper we review the current experimental and
theoretical situation of the description
of non-Fermi liquid behavior (NFL) in U and Ce intermetallics.
We focus on the magnetic and thermodynamic properties. 
We also discuss a recent theoretical interpretation of this behavior in
terms of Griffiths-McCoy singularities close to the magnetic
quantum critical point (QCP). We show how an effective Hamiltonian
which contains both the RKKY coupling and the Kondo interaction
can be written after high energy degrees of freedom away from the Fermi
surface are traced out. We argue that dissipation due
to particle-hole excitations close to the Fermi surface 
is a relevant
perturbation at low temperatures and we estimate the crossover
temperature $T^*$ above which power law behavior in specific heat
and magnetic response occurs ($C_V/T \sim \chi(T) \propto T^{-1+\lambda}$
with $\lambda<1$). Below $T^*$ a new regime dominated by
dissipation is found and deviations from power law behavior
are expected. 
\end{abstract}

\newpage

\narrowtext

\section{Introduction}

The basis for the study of metals was set by Landau almost 50
years ago in his studies of He$^3$. 
During all these years the Landau theory
has been a paradigm used to explain the experimental
behavior and electronic properties of quantum Fermi liquids \cite{Baym,pines}.
Initially the theory appeared as a phenomenological framework with a few
parameters fixed by experiments. The presence of
unknown parameters reflected, at that time, the lack
of a microscopic theory.
However, it was an extraordinary and necessary
first step. Landau himself also established the route for
the microscopic explanation for the validity of the theory. The Landau
theory became the main tool for the study of the effects of correlations
in electronic systems, and its foundation was eventually established on
microscopic grounds using field theoretic methods
\cite{abrikosov,nozieres}. The theory is supposed to hold at
temperatures much lower than the Fermi temperature of the system.
It is based mostly on the assumption that the interaction
among electrons is short ranged (due to screening) and they
are such that a perturbative expansion in the interaction converges.
Thus there is a one-to-one correspondence between the
interacting system of electrons and a {\it weakly} interacting
system of quasiparticles. Moreover, the physics of the
quasiparticles is completely determined by the Fermi surface.

As consequences of Landau's theory the thermodynamic and response functions
of the electron fluid are smooth functions of the temperature.
One would have for instance a temperature
independent Pauli susceptibility, $\chi(T) \propto$ constant,
a temperature independent specific heat coefficient, $\gamma(T) =
C_V/T \propto$ constant, and a Korringa law for the NMR relaxation rate,
$1/(T_1 T) \propto$ constant. 
Furthermore,
at low temperatures one expects the electronic resistivity to
behave like 
$\rho(T) = \rho_0 + A T^2$, 
where $\rho_0$ is the resistivity due to impurities and
$A$ is a coefficient which comes from three different sources: electron-electron
Umklapp processes \cite{ziman1}, electron-electron interactions mediated by
phonons \cite{mac}, and the inelastic scattering of electrons
by impurities \cite{koshino}. These predictions have been confirmed
in a wide class of metals and are considered the trademark of
Fermi liquid behavior.

Violations of Fermi liquid behavior have been expected for a long
time in the context of one-dimensional conductors \cite{1d} 
due to strong restrictions in phase space for electron-electron
scattering. 
It turns out, however, that it is 
very hard to experimentally observe NFL behavior
in one-dimensional systems.
In the case of organic conductors \cite{jerome}, which are
considered the prototype of one-dimensional metals, there is
always a
crossover to higher dimensional behavior (that is, Fermi liquid
behavior) due to coupling between chains at low temperatures.
The only clear observation of NFL behavior in low dimensional
systems appears in the 
very special case of the edge states of quantum Hall bars 
\cite{stone} where NFL behavior is due to the Landau level degeneracy.
Indeed, there is a strong controversy about the possibility of
NFL behavior in dimensions higher than one. The subject
was raised by Anderson in the context of high temperature superconductors
\cite{anderson}. Although the possibility of NFL behavior
in 2 dimensions has not been discarded there are today strong
arguments against it \cite{engel,bosoeu,bosoeles,bosoko,castel}.
In 3 dimensions Landau's Fermi liquid theory is assumed to
be the correct starting point. We stress
that even in the presence of disorder Fermi liquid theory
should be valid in 3 D (at least when the disorder
is weak enough to be treated in perturbation theory)\cite{bill}.
Thus, it is indeed very surprising
that for such a broad class of U and Ce
alloys (which clearly show
three-dimensional behavior) deviations from Landau's theory
are so abundant. Actually, there are nowadays so many examples
of alloys presenting NFL behavior that the discovery of a new
compound which exhibits such a behavior is not a surprise.

We organize this paper as follows: in the next section we give a
brief
overview of the theoretical and experimental situation
on the problem of NFL behavior in U and Ce intermetallics.
We apologize in advance to any whose work we have unintentionally
left out. We concentrate on the situation of the thermodynamic
and magnetic response in these systems and leave the important
problem of transport for a later publication. In
Section \ref{grifinsu} we discuss the problem of Griffiths-McCoy
singularities in insulating magnets; in Section \ref{klat}
we discuss the Kondo lattice problem and show how the RKKY
interaction and the Kondo effect appear at the Hamiltonian
level when high energy degrees of freedom are eliminated
from the Hamiltonian; in Section \ref{dissipation}
we discuss the differences between the insulating case and
the metallic case for the formation of Griffiths-McCoy singularities;
finally Section \ref{conclusions} contains our conclusions.

\section{Overview of the NFL behavior in U and Ce intermetallics}
\label{overview}

The systems we are considering in this paper are metallic
alloys of rare earths or actinides which can be classified
as (1) {\it Kondo hole} systems, in which 
the rare earth or actinide (R) is substituted 
by a non-magnetic metallic atom (M) with a
chemical formula R$_{1-x}$M$_{x}$ (a typical example is
U$_{1-x}$Th$_x$Pd$_2$Al$_3$); (2){\it Ligand systems}, where
one of the metallic atoms (M1) is replaced by another (M2) but the
rare earths or actinides are not touched and thus have the formula 
R(M1)$_{1-x}$(M2)$_x$ (as, for instance, UCu$_{5-x}$Pd$_x$).
Often these alloys order magnetically
at $x=0$ (ordered Kondo lattices) and long range order is lost
at some $x=x^*$ as shown in Fig.\ref{phdi}. For $x>x^*$ the
ordered state is replaced by a metallic state which
shows physical properties which deviate strongly from the
predictions of Fermi liquid theory. This state is called
non-Fermi liquid state (NFL). 

In NFL systems it is usually observed that even in the
paramagnetic phase the specific heat coefficient and the magnetic 
susceptibility
do not saturate as expected from the Landau scenario. 
The theoretical reason for this anomalous behavior is still
not completely understood; many different theories have
been proposed and the subject is very controversial.
One possible
reason for singular behavior in the thermodynamic and response
functions of the system is due to closeness of these systems to
long-range order. The idea that a quantum critical point (QCP) 
could be responsible for
NFL behavior was proposed by Hertz \cite{hertz} and later
extended by Millis and Continentino \cite{millis}. Indeed,
there is strong evidence that QCP physics is responsible for
NFL behavior in CeCu$_{6-x}$Au$_x$, where NFL behavior can
be fine tuned via magnetic fields or pressure to the QCP \cite{von}.
It turns out, however, that even for this compound there is controversy
about the correct description of the QCP \cite{piers,2dcecuau}.
Another system recently studied which seems to be in this category
is CeNi$_2$Ge$_2$, which has been shown to have a minimum amount
of disorder \cite{steglich}. Hertz also studied the problem of 
disorder in a XY magnet and found disorder to be a relevant
perturbation \cite{hertzxy}. In this context Hertz conjectured
that disorder could lead to clustering of magnetic moments.
More recently it has been shown that quenched disorder has
a strong effect on the properties of quantum antiferromagnets
and leads to very unconventional critical behavior \cite{vojta}. 

In all the cases studied so far the data for the susceptibility and specific heat have
been fitted to weak power laws or logarithmic functions \cite{staba}.
The resistivity of the systems discussed here can be 
fitted with $\rho(T) = \rho_0 + A T^{\alpha}$ where $\alpha < 2$. 
Neutron scattering experiments
in UCu$_{5-x}$Pd$_x$ \cite{ucupdn} show that the imaginary
part of the frequency dependent susceptibility, $\Im (\chi(\omega))$,
has power law behavior, that is, $\Im (\chi(\omega)) \propto \omega^{1-\lambda}$
with $\lambda \approx 0.7$, over a wide range of frequencies 
(for a Fermi liquid one expects $\lambda=1$). Moreover, consistent
with this behavior the static magnetic susceptibility seems
to diverge with $T^{-1+\lambda}$ at low temperatures \cite{ucupdn}.
What is interesting about UCu$_{5-x}$Pd$_x$ is that it has been
shown in recent EXAFS experiments that this compound has a
large amount of disorder \cite{ucupdexaf} consistent
with early NMR and $\mu$SR experiments \cite{ucupdnmr}. 
Even in stoichiometric systems like CeAl$_3$ there is
evidence of spatial inhomogeneity \cite{ott}. In CePd$_2$Al$_3$
it has been shown that while polycrystalline samples show 
a magnetic phase transition at finite temperatures the critical
temperature is driven to zero in single crystals
due to internal stresses which suppress the moment formation \cite{mentink}.
Moreover, UCu$_4$Pd is supposed to be exactly at the QCP for
{\it antiferromagnetic} order. All these properties
have also been seen in a similar alloys such as  UCu$_{5-x}$Al$_x$
and UCu$_{5-x}$Ag$_x$ \cite{ucual}. 
Another system which is also close to a magnetic
order is U$_{1-x}$Y$_x$Pd$_3$ which shows {\it spin glass} order
\cite{uypd}. It was the study of this system which lead
Andraka and Tsvelik to propose that the NFL behavior observed in
this compound was due to the spin glass transition at the QCP
\cite{tsvelik}. This point of view has also been explored by other
researchers in the field \cite{sgsub}. Another interesting example
where NFL behavior happens close to a QCP is in the system 
U$_{1-x}$Th$_x$Cu$_2$Si$_2$ which shows a {\it ferromagnetic}
QCP \cite{uthcusi}. Thus, NFL behavior has been observed in
systems with very different types of magnetic ordering. Indeed,
the magnetic behavior in these systems is very rich. Very recent
frequency dependent susceptibility measurements have found
signs of super-paramagnetism (that is, cluster physics \cite{mydosh}) close
to the QCP of UCu$_{5-x}$Pd$_x$ \cite{ucupdchi}. Fluctuating magnetic
moments were also found in $\mu$SR experiments in Ce(Ru$_{1-x}$Rh$_x$)$_2$Si$_2$
close to the QCP \cite{miyako}. Actually, pressure 
experiments in the same compound have revealed
the importance of disorder for the appearance of NFL behavior \cite{graf}.
The same type of physics is found in very complex systems such as
U$_{3-x}$Ni$_3$Sn$_{4-y}$ \cite{lance} and  
even in stoichiometric alloys like Yb$_2$Ni$_2$Al where a coexistence of
magnetic and paramagnetic phases have been observed \cite{winkel}. 
The phenomena observed here are indeed very similar to those observed in
simpler magnetic alloys close to a magnetic phase transition as has
been shown in recent experiments in Ni$_x$Pd$_{1-x}$ close to
a ferromagnetic QCP \cite{niklas}. Indeed, the phenomena of clustering, superparamagnetism, 
and magnetic fluctuations has been discussed long ago in simple alloys
such as Ni$_x$Cu$_{1-x}$ both theoretically and experimentally \cite{nicu}.
As in their f-electron counterparts, the specific heat and magnetic
susceptibility of these systems deviate strongly from Fermi liquid behavior in a relatively
broad region around the quantum critical point.
Indeed, the phenomenon of NFL behavior is
not limited to f-electron systems but is also very common in d-electron
compounds \cite{julian}.

Another source of NFL behavior is of single impurity nature.
Single impurity approaches for the NFL problem are very
important because the Kondo effect \cite{zener,kondo} is known to happen
in the dilute limit of Kondo hole systems (e.g., in U$_{1-x}$Th$_x$Pd$_3$
for $x \approx 1$) and has been suggested as the source of
{\it heavy fermion} behavior \cite{hewson,greg} in undiluted ligand systems (e.g., CeAl$_3$)
\cite{klattice,sigrist}. 
The Kondo effect is probably one of the most studied problems in many-body
theory. It has been understood from many different points
of view, from renormalization group (RG) calculations \cite{wilson}
to the exact analytic solution \cite{bethe} and from the conformal field theory
point of view \cite{affleck}. In the anisotropic case the Kondo Hamiltonian
can be mapped via bosonization into
the dissipative two level system (DTLS) \cite{twolevel,twolevelother}. 

The mapping between
the Kondo problem and DTLS was developed in order to understand the
problem of a quantum phase transition in the DTLS \cite{bray,sudip}
and was believed to be valid close to the QCP. Nowadays, extensive
numerical simulations have shown that the mapping is valid
over most of the parameter space \cite{costi}. 
In the single channel Kondo problem
up and down spin electrons spin-flip scatter against the magnetic
moment. Nozi\`eres and Blandin proposed that 
when the number of scattering channels is increased one can obtain a
local NFL ground state \cite{over}. This is the so-called {\it multichannel
Kondo effect}. In 1986 D.~Cox proposed an elegant 
mechanism for NFL behavior in
U systems based on a multichannel effect of quadrupolar origin \cite{cox}. 
The same mechanism has been studied in the context of
the Kondo lattice \cite{coxagain}. There is still 
controversy about the applicability of the quadrupolar Kondo effect 
to the compounds we are discussing \cite{agcox}. The quadrupolar
Kondo effect requires a non-magnetic $\Gamma_3$ ground state which
has been confirmed to exist in PrInAg$_2$ \cite{galera}. It turns
out that the experimental situation in this compound is far from clear:
while the magnetic susceptibility seems to show NFL behavior
the specific heat is well described by Fermi liquid theory \cite{ward}. 
Moreover, the exponents
predicted by multichannel effects are not consistent with the experimental
data in many compounds. Another source of NFL
behavior based on single impurity physics is the so-called {\it Kondo disorder}
approach in which it is assumed that due to the intrinsic disorder in the
Kondo lattice there is a broad distribution of Kondo temperatures going down
to a vanishing one \cite{ucupdnmr}. This kind of approach has
been applied to UCu$_{5-x}$Pd$_x$ \cite{miranda}. Naturally,
the main criticism to the single impurity approaches is that the systems
where NFL behavior is observed are concentrated. Furthermore, NFL behavior
usually occurs close to a QCP where interactions among
the moments and tendency to magnetic ordering is very important. 
Thus, the problem of single impurity versus QCP physics as
a source of NFL behavior remains a
very controversial one and debate among researchers is still in progress.

In the next sections we discuss a possible explanation for the NFL
observed in these alloys which is due to Griffiths-McCoy singularities
close to the QCP point \cite{grifeu}. The origin of these singularities
is due to the competition between the RKKY interaction, which leads
to magnetic order, and the Kondo effect, which leads to magnetic quenching
in the presence of disorder due to alloying. 

\section{Griffiths-McCoy singularities}
\label{grifinsu}

The simplest way to understand the nature of Griffiths singularities
is to imagine the dilution of a magnetic lattice by non-magnetic
atoms. Long range order is lost at percolation threshold when the
last infinite cluster of magnetic moments seize to exist. Above the threshold
the system is composed of finite clusters of magnetic atoms. Griffiths
showed that when a magnetic field is applied to the percolating lattice 
there is a non-analytic contribution of the 
clusters to the free energy \cite{grif}. This contribution comes
from {\it rare} large clusters. The classical
problem was studied in great detail by many researchers in the 70's
\cite{grifother}. An important special model related with the problem
of Griffiths singularities was proposed by McCoy and Wu \cite{mccoy}
and studied more recently by Shankar and Murthy \cite{shankar}. 
The McCoy-Wu model is a rectangular Ising model with disorder
in only one direction. The importance of this model relies on
the fact that it is the only known exactly solvable model with disorder.
Moreover, it was shown that in this model while the system orders magnetically
at some temperature $T_c$ the magnetic susceptibility diverges before
the system reaches $T_c$. This strange behavior is again due to
Griffiths singularities.
Although classical Griffiths singularities are rather weak (and for
a long time researchers believed only the singularity coming from
the infinite cluster would be observable experimentally) there is
recent experimental evidence for their existence in some Ising magnets
\cite{grifexp,belanger}.

It turns out that from a statistical mechanical point of view
a classical 2 D Ising problem is equivalent to
a 1 D quantum Ising model at zero temperature. Thus, the McCoy-Wu problem maps,
at zero temperature, into the random transverse field Ising chain.
The random transverse field Ising model was studied in great detail
by D.~Fisher, who was able to calculate asymptotically exact
expressions for many physical quantities close to the QCP
\cite{dfisher}. The random transverse field Ising model has also been
extensively 
studied analytically \cite{senthil,huse} and numerically \cite{tfinum}
in one and higher dimensions and evidence for Griffiths-like singularities
was obtained. In particular, Thill and Huse proposed a {\it quantum
droplet model} for the problem where the magnetic cluster is treated
like a single degree of freedom (or two level system). This kind
of treatment agrees very well with the exact 1D calculation.
It has been shown that the same type of singularities can
also happen in more complicated random magnetic systems such as in the
XY model \cite{ross}.
One of the main characteristics of the Griffiths-McCoy
singularities is the divergence of the physical quantities at zero
temperature with non-universal power laws. The
magnetic susceptibility, for instance, behaves like $\chi(T) \propto T^{-1+\lambda}$ 
and diverges in the {\it paramagnetic phase} if $\lambda<1$,  
and the nonlinear susceptibility diverges with an even stronger
power law, $\chi_{nl}(T) \propto T^{-3+\lambda}$ 
(indeed it has been shown that for
the pure Ising model $\lambda \to 0$ at the QCP \cite{senthil}). 
Experimental evidence for a divergent non-linear susceptibility was found in
the compound U$_{1-x}$Th$_x$Be$_{13}$ \cite{aliev}. 

That Griffiths singularities should be important in some single band
correlated systems was proposed by Bhatt and Fisher \cite{bhatt} and extended more recently 
by Sachdev \cite{grifsach}. A similar type of phenomenon happens
also in magnetically doped semiconductors where a {\it
singlet phase} was proposed by Bhatt and Lee \cite{lee}.
It was shown that for a single band Hubbard model
in a disordered environment the magnetic properties could be explained in
terms of a {\it quantum spin glass} ground state \cite{sachmore}.
We have proposed recently that the
Kondo lattice problem is in
the same class of problems as the random transverse Ising model \cite{grifeu}.
Dipolar interactions are too small to account
for the ordering temperature in these systems (which range from $100$ K down to
$10$ K in the pure compounds), and the direct exchange between f orbitals is very weak
since the spatial extent of the f orbitals is small.
As is well-known the magnetism in Kondo
lattices comes from the localized f-moments which are weakly
hybridized with the conduction band. In the pure compound (say,
UPd$_2$Al$_3$ or UCu$_5$) the moments interact with each other via
the RKKY interaction which is propagated by the conduction band \cite{rudkit}
and in the presence of disorder and spin-orbit effects the RKKY interaction
becomes short ranged \cite{elihu}.
Moreover, most of the heavy fermion alloys are magnetically
anisotropic because of crystal field effects or spin-orbit
coupling which are known to be very important in these systems.
In this case the magnetic phase diagram can be very rich since an 
anisotropic spin exchange interaction, or Dzyaloshinsky-Moriya (DM) exchange 
interaction \cite{dzmo}, is generated. These anisotropies have been observed 
long ago in alloys of rare earths of the form 
R-CrO$_3$ \cite{yosida,koeler} where R is a rare earth.

In a Kondo hole system (say, U$_{1-x}$Th$_x$Pd$_2$Al$_3$) the
destruction of magnetism occurs mainly by the dilution
of the magnetic lattice and the QCP is the percolation threshold
for the lattice. In the ligand systems (like, say, in UCu$_{5-x}$Pd$_x$)
the dilution is a more subtle effect because the magnetic atoms
remain on the lattice. 
In order to understand how dilution quenches
a magnetic moment in a ligand system one has to look at what happens in the related
heavy fermion materials which do not show long range order. In
heavy fermions the magnetic moments are quenched by the Kondo effect
\cite{klattice} as described, for instance, in the {\it dynamical
mean field} (or $d \to \infty$) theories of the Anderson lattice \cite{dinf}.
 Thus, it is reasonable to assume that in ligand systems
the effect of doping is to affect {\it locally} the hybridization
between localized moments and conduction electrons. 
This naturally leads us to the picture proposed long ago by Doniach
\cite{doniach}, namely, there are two relevant energy scales in
the problem: the Kondo temperature of the moment $T_K$ which is
an exponential function of the exchange $J$ between local moments
and the itinerant electrons ($J \propto V^2$ where $V$ is the hybridization
matrix element and $T_K \propto \exp[-1/(N(0) J)]$ where $N(0)$
is the density of states at the Fermi surface) and the RKKY temperature
scale, $T_{RKKY}$, associated with magnetic ordering of the moments
which scales with $J^2$. As shown in Fig.\ref{doni} there is a critical value
of $J$ (say, $J_c$) for which these two energy scales become of the
same order of magnitude. For $J<J_c$ we have $T_K < T_{RKKY}$ and therefore,
as the temperature is lowered the system orders magnetically before
the moment is quenched. If $J>J_c$, that is, if $T_K > T_{RKKY}$, as the
temperature is lowered the moment is quenched before it has the
chance to order. Although this picture is quite naive it has been
confirmed in mean field theories \cite{klmf,coqblin}, numerical calculations \cite{klnum} 
and pressure experiments in ordered\cite{klexp}  and disordered Kondo lattices\cite{klexpalloy} . 
In the absence of disorder this competition leads to a
finite ordering temperature $T_N$ which vanishes at a QCP at $J_c$ as shown in Fig.\ref{doni}.
Moreover, the competition appears explicitly in a simple problem
of two interacting moments in the presence of a Fermi sea which is
the {\it two impurity} Kondo problem \cite{twoimp}. Numerical
works on this problem have confirmed the theoretical expectations 
\cite{twoimpnum}.
Thus, in our scenario long range order is lost in a ligand system
by the {\it local} quenching of the magnetic moments. Again one
has to deal with a {\it quantum} percolation problem. The QCP
is again the percolation point of the magnetic lattice. Away from
the QCP only magnetic clusters can exist. As one further dopes the
system away from the magnetic phase one eventually finds a heavy
fermion ground state where all the moments are quenched \cite{klattice} (unless,
of course, there is a structural or another magnetic phase transition
in the intermediate region).

In order to describe mathematically the competition between the Kondo
effect and the RKKY interaction we may start with the mean field description
of the ordered phase (this is a good first approximation since there 
is true long range order in the system at finite temperatures and
quantum fluctuations are small). On the one hand the conduction
electron band is renormalized by the average field created by the 
ordered magnetic moments.
Unless there are commensuration effects between the magnetic
ordering vector and the Fermi momentum the electrons remain gapless
(such commensuration effects are very unlikely in these systems with
complex unit cells). On the other hand the conduction electrons
produce an effective medium for the propagation of the RKKY interaction.
Suppose one dilutes slightly the ordered state by changing a local
exchange constant $J$ to a value much larger than the average. In this
case, like in the Doniach argument, the local moment instead of participating
in the collective magnetic state will prefer to form a local singlet.
That is, one has again a simple Kondo effect with a renormalized conduction
band. The magnetization of the system has to drop. Mathematically
this quenching of the magnetic moment can be described in terms of the
anisotropic Kondo problem. Actually, as we discussed,
the magnetism in these systems is anisotropic and therefore the Kondo
effect does not have SU(2) symmetry. Indeed there are very recent
inelastic neutron scattering experiments in Ce$_{1-x}$La$_x$Al$_3$ which 
show evidence for anisotropic Kondo behavior \cite{osborn}. Therefore,
as mentioned previously the Kondo effect can be mapped into the
DTLS. As we discuss in the next section 
the XY part of the Kondo problem becomes a transverse
field - the origin of this term can be thought as a transverse
magnetic field applied by the electron spin on the magnetic moment -
and the Ising component of the Kondo exchange describes the coupling
of the magnetic moment to a heat bath - which represents the fact that
each time the magnetic moment flips it produces particle-hole excitations
at the Fermi surface. It turns out that the coupling to the heat bath
becomes small in the limit of large anisotropy and {\it in zeroth order}
can be disregarded. Thus, we have argued \cite{grifeu}
that in this extreme limit
the Kondo lattice problem maps into the random transverse field
Ising model which, as we said previously, has been shown to present
Griffiths-McCoy singularities with power law behavior at low temperatures.
Recent experiments have shown
that power law behavior is consistent with measurements of
magnetic susceptibility and specific heat in these systems \cite{marcio}. In particular,
for the Griffiths phase one has $\chi(T) \sim \gamma(T) \propto T^{-1+\lambda}$
with $\lambda<1$. In the next section we show that the residual coupling
with the particle-hole bath changes the behavior of the response
functions below a certain energy scale. 

The Griffiths phase picture has been very successful in describing the power law
behavior of the physical quantities in some of the systems mentioned above, especially,
UCu$_{5-x}$Pd$_x$ where structural disorder was clearly measured \cite{ucupdnmr}
and the exponents measured from specific heat and susceptibility agree well
with each other ($\lambda \approx 0.72$) \cite{marcio}. 
In other alloys such as U$_{1-x}$Th$_x$Pd$_2$Al$_3$
($\lambda \approx 0.8$ from specific heat and
$\lambda \approx 0.63$ from susceptibility data) \cite{marcio} or 
Ce(Pd$_{1-x}$Ni$_x$)$_2$Ge$_2$ ($\lambda \approx 0.7$ from specific heat and
$\lambda \approx 0.84$ from susceptibility data) \cite{knebel} the agreement
between exponents is not as good (although, we should stress, power law behavior
is clearly observed). Moreover, in systems like U$_{0.2}$Y$_{0.8}$Pd$_3$
the divergence seems to be stronger than power law or logarithm \cite{marcio}.
One possible explanation for these stronger divergences might be related
with the possibility that the disorder in these systems is correlated and not random. In
this case one expects stronger divergences \cite{rieger}. 
Furthermore, in some of the systems described 
above there may be a return to Fermi liquid behavior at very low temperatures (for instance,
in UCu$_4$Pd is evidence of thatbelow $0.1$ K \cite{ucupdchi} while in CeRhRuSi$_2$ 
it seems to occur below $1$ K \cite{graf}). 
We argue below that some of these problems can be resolved if the metallic character
of the electronic environment is taken into consideration. 

\section{Griffiths singularities and the Kondo lattice problem}
\label{klat}

It is intuitively obvious that the Doniach argument lead to an inhomogeneous
behavior when it is taken locally instead of globally. The main problem here
is how this argument can be tested at the Hamiltonian level. In this section
we show how this can be accomplished for the Kondo lattice model from the
renormalization group point of view \cite{shankarg}. 

It is imperative in the context of the systems discussed in this paper
to take into account spin-orbit effects since these are very important
for f-electron magnetism. In this case the exchange between local moments
and conduction electrons is not isotropic in spin space and can
be generally written as \cite{next}:
\begin{eqnarray}
H = \sum_{{\bf k},\kappa} \epsilon_{\kappa}({\bf k}) 
c^{\dag}_{{\bf k},\kappa} c_{{\bf k},\kappa}
+ \sum_{i} \sum_{a,b,\kappa,\kappa'}
J_{a,b} (i) S^a(i)  
c^{\dag}_{\kappa}(i) \tau_{\kappa,\kappa'}^b c_{\kappa'}(i)
\label{kondoarray}
\end{eqnarray}
where $\kappa=1,2$ labels the spin states in the diagonal basis,
$J_{a,b}$ are the effective exchange constants between the
localized spins, $S^a(i)$, and the conduction electron
spin, $\sum_{\kappa,\kappa'}c^{\dag}_{\kappa}(i) \tau_{\kappa,\kappa'}^b c_{\kappa'}(i)$. In the simplest case of uniaxial symmetry (which is going
to be discussed throughout this paper) one has $J_{a,b} = J_a \delta_{a,b}$
where $J_z > J_x=J_y = J_{\perp}$.

The main problem in studying the competition of RKKY and Kondo effect in the
Hamiltonian (\ref{kondoarray}) is related with the fact that both
the RKKY and the Kondo effect have origin on the same magnetic coupling
between spins and electrons. What allows us to treat this problem is the
fact that the RKKY interaction is perturbative in $J/E_F$ while the
Kondo effect is not. Moreover, the RKKY interaction depends on electronic
states deep inside the Fermi sea while the Kondo effect is a Fermi
surface effect. Thus, it seems to be possible to use perturbative
renormalization group approach to treat the RKKY interaction while for
the Kondo effect one needs to do a better job. This kind of treatment
was proposed recently in the context of the two impurity Kondo problem
\cite{baraf}.

We will consider, for simplicity, the case where the Fermi surface for the
electrons is spherical (non-nested, non-spherical Fermi surfaces can be treated
in an analogous way). The local electron operator can
be written in momentum space as
\begin{eqnarray}
c_{\kappa}({\bf r}) = \frac{1}{\sqrt{N}} \sum_{{\bf k}} e^{i {\bf k} \cdot {\bf r}} \, c_{{\bf k},\kappa} \, .
\label{cki}
\end{eqnarray}
We now separate the states in momentum space into three different regions of energy
as shown in Fig.\ref{fs}, namely,
$\Omega_0$ where $k_{F}-\Lambda<k<k_{F}+\Lambda$; 
$\Omega_1$ where $k<k_{F}-\Lambda$; and $\Omega_2$ where $k>k_{F}+\Lambda$ where $\Lambda$ is
an arbitrary cut-off. Observe that in this case the sum in (\ref{cki}) can now also
be split into these three different regions. The problem we want to address
is how the states in region $\Omega_0$ close to the Fermi surface renormalize as one traces out high energy
degrees of freedom which are present in regions $\Omega_1$ and $\Omega_2$. We can perform this calculation
perturbatively in $J/E_F$. For that purpose it is more convenient to use a path
integral representation for the problem and write the quantum partition function as
\begin{eqnarray}
Z= \int D{\bf S}(n,t) D\bar{\psi}({\bf r},t) D\psi({\bf r},t) \exp\left\{
\frac{i}{\hbar} {\cal S}[{\bf S},\bar{\psi},\psi] \right\}
\label{zspsi}
\end{eqnarray}
in terms of Grassman variables $\bar{\psi}$ and $\psi$ and
where the path integral over the localized spins also contains the constraint that
${\bf S}^2(n,t)=S(S+1)$. The quantum action in (\ref{zspsi}) can be separated into
three different pieces, ${\cal S}={\cal S}_0[{\bf S}]+{\cal S}_0[\bar{\psi},\psi]
+{\cal S}_I[{\bf S},\bar{\psi},\psi]$, where ${\cal S}_0[{\bf S}]$ is
the free actions of the spins (which can be written, for instance, in terms of
spin coherent states \cite{fradkin}), 
\begin{eqnarray}
{\cal S}_0[\bar{\psi},\psi] =  
\sum_{\alpha,\gamma} \int d\omega \int d{\bf k} 
\bar{\psi}_{\alpha}({\bf k},\omega) 
\left(\omega + \mu - \epsilon_{\alpha}({\bf k}) \right) \delta_{\alpha,\gamma} \psi_{\gamma}({\bf k},\omega)
\label{s0psi}
\end{eqnarray}
is the free action for the conduction electrons and
\begin{eqnarray}
{\cal S}_I[{\bf S},\bar{\psi},\psi] = \sum_{\alpha,\gamma}  
\sum_{n,a,b} \int dt \, J_{a,b}(n) S_a(n,t) \tau^b_{\alpha,\gamma} \bar{\psi}_{\alpha}(n,t) \psi_{\gamma}(n,t)
\label{interaction}
\end{eqnarray} 
is the exchange interaction between conduction electrons and localized moments.

We can now split the Grassman fields into the momentum shells
defined above, that is, we rewrite the path integral as
\begin{eqnarray}
Z= \int D{\bf S}(n,t) \prod_{i=0}^2 D\bar{\psi}_i({\bf r},t) D\psi_i({\bf r},t) \exp\left\{
\frac{i}{\hbar} {\cal S}[{\bf S},\{\bar{\psi}_i,\psi_i\}] \right\}
\end{eqnarray}
where the index $0,1,2$ refers to the degrees of freedom which reside in the momentum
regions $\Omega_0$, $\Omega_1$ and $\Omega_2$, respectively. In this case the action 
of the problem can be rewritten as ${\cal S} = {\cal S}_0[{\bf S}] + \sum_{i=0}^2 {\cal S}_0[\bar{\psi}_i,\psi_i]
+ {\cal S}_I[{\bf S},\bar{\psi},\psi]$. Notice the free part of the electron action is just a sum
of three terms (essentially by definition since the non-interacting problem is diagonal in momentum
space). Moreover, the exchange part mixes electrons in all three regions defined above:
\begin{eqnarray}
{\cal S}_I = 
\sum_n \sum_{i,i'=0}^2 \int dt \, J_{a,b}({\bf r}_n) S_a({\bf r}_n,t) \tau^b_{\alpha,\gamma}
\bar{\psi}_{\alpha,i}({\bf r}_n,t) \psi_{\gamma,i'}({\bf r}_n,t) \, .
\label{lsplit}
\end{eqnarray}

Since we are interested only on the physics close to the Fermi surface we trace out
the fast electronic modes in the regions $\Omega_1$ and $\Omega_2$ assuming that
$J_{a,b}\ll\mu$. In this case, as we show elsewhere \cite{next}, besides
the renormalization of the parameters in free action of the electrons in the
region $\Omega_0$, we get the RKKY interaction between localized moments, that is,
the effective action of the problem becomes:
\begin{eqnarray}
{\cal S}_{eff}[{\bf S},\bar{\psi}_0,\psi_0] &=& {\cal S}_0[\bar{\psi}_0,\psi_0] + 
\sum_n \sum_{\alpha,\gamma,a,b} \int dt \, J^R_{a,b}({\bf r}_n) S_a({\bf r}_n,t) \tau^b_{\alpha,\gamma}
\bar{\psi}_{\alpha,0}({\bf r}_n,t) \psi_{\gamma,0}({\bf r}_n,t) 
\nonumber
\\
&+& {\cal S}_0[{\bf S}] + \sum_{n,m,a,b} \int dt \, \Gamma^R_{a,b}({\bf r}_n-{\bf r}_m) S_a({\bf r}_n,t) S_b({\bf r}_m,t)
\label{almosthere}
\end{eqnarray}
where $\Gamma^R_{a,b}({\bf r}_n-{\bf r}_m)$ is the cut-off dependent
RKKY interaction between the local moments and $J^R_{a,b}(n)$ is the {\it Kondo}
electron-spin coupling renormalized by the high energy degrees of freedom.
The renormalization can be calculated
order by order in perturbation theory \cite{next}. We observe further that the perturbation
theory here is well behaved and there are no infrared singularities in the
perturbative expansion. Thus, the limit of $\Lambda \to 0$ is well-defined.
In this limit $\Gamma_{a,b}({\bf r}_n-{\bf r}_m,\Lambda \to 0)$ becomes
the usual RKKY interaction one would calculate by tracing all the energy
shells of the problem. Observe that there are no retardation effects
in tracing this high energy degrees of freedom since they are much faster
than the electrons close to the Fermi surface and therefore adapt adiabatically
to their motion (the situation here is somewhat similar to the one in the
Born-Oppenheimer approximation where the ions are much slower than the electrons
and therefore can only renormalize the coupling constants).
Hamiltonian (\ref{almosthere}) is the basic starting
point of our discussion and contains the basic elements for the discussion
of magnetic order in the system. 
As discussed in the Doniach's argument, 
the RKKY interaction tends
to order the magnetic moments while the Kondo coupling tends to quench it.  
It is the interplay of these two interaction which leads to the physics
we discuss here.
The way this quenching occurs its fundamental for the understanding of 
physics of this problem and it is discussed in the next section.

\section{Dissipation in metallic magnetic alloys}
\label{dissipation}

In order to understand our line of argument it is important to study a very simple
case of (\ref{almosthere}) where RKKY interactions are not present, that is,
the single impurity Kondo problem. As we have said previously this problem is
well understood and here we will just quote a few of the important results
for our discussion. Notice that the single impurity Kondo effect should
occur in the case of Kondo hole systems when $x \ll 1$ or in moment
quenching in ordered ligand systems when $x \approx 1$.

For the single impurity the mathematical description simplifies
greatly because we just have to solve a scattering problem in terms
of in-coming and out-going waves. Thus, the problem is effectively
one dimensional with a boundary condition at the impurity position \cite{twolevel,twolevelother}. In this case we can use the technique of bosonization
to understand the basic physics. First of all we can
show that the renormalization of the Ising component of the Kondo
interaction (\ref{almosthere}) is given in terms of the phase shift
$\delta$ of the electrons by the impurity:
\begin{eqnarray}
J_z^R = 8 v_F \delta(J_z)
\label{jzr}
\end{eqnarray}
where $v_F$ is the Fermi velocity and
\begin{eqnarray}
\delta(J_z) = \arctan\left(\frac{\pi N(0) J_z}{4}\right) \, .
\label{djz}
\end{eqnarray}
Since we are treating states very close to the Fermi surface 
we linearize the electron dispersion close to the Fermi surface:
\begin{eqnarray}
E_{{\bf k}} = E_F + v_{F} (|{\bf k}|-k_{F}) \, ,
\label{linear}
\end{eqnarray}
in which case the the conduction
band Hamiltonian is written as
\begin{eqnarray}
H_{C} = \sum_{p,\sigma} v_{F} p c^{\dag}_{p,\sigma} c_{p,\sigma}
\label{hc}
\end{eqnarray}
where $c_{p,\sigma}$ creates an electron with spin $\sigma$,
momentum $|{\bf k}| = p + k_F$ and angular momentum $l=0$.
Thus, in writing (\ref{hc}) we have reduced the problem to an
effective one-dimensional problem. We introduce 
right, $R$, and left, $L$, moving electron operators
\begin{eqnarray}
\psi_{R(L),\sigma}(x) = \frac{1}{\sqrt{2 \pi}} \int_{-\Lambda}^{\Lambda}
dk c_{k \pm k_{F,\sigma},\sigma} e^{i k x}
\end{eqnarray}
which are used to express the electron operator as
\begin{eqnarray}
\psi_{\sigma}(x) = \psi_{R,\sigma}(x) e^{i k_{F,\sigma} x}
+ \psi_{L,\sigma}(x) e^{-i k_{F,\sigma} x} \, .
\end{eqnarray}
In any impurity problem the right and left moving operators
produce a redundant description of the problem since they
are actually equivalent to in-coming or out-going waves out of
the impurity. Therefore we have two options: either we work with
right and left movers in half of the line or we work in the full
line but we impose the condition  $\psi_{R,\sigma}(x) = \psi_{L,\sigma}(-x)$.
We will use the last option. Thus, from now on we drop the symbol $R$
from the problem and work with left movers only. 
The left mover fermion can be bosonized as 
\begin{eqnarray}
\psi_{\sigma}(x) = \frac{K_{\sigma}}{\sqrt{2 \pi a}} 
e^{i \Phi_{\sigma}(x)}
\label{mandel}
\end{eqnarray}
where
\begin{eqnarray}
\Phi_{\sigma}(x) = \sum_{p>0} \sqrt{\frac{\pi}{p L}}
\left((b_p + \sigma a_p) e^{i \sigma p x} - h.c.\right)
\end{eqnarray}
and $K_{\sigma}$ is a factor which preserves
the correct commutation relations between electrons, that is,
$\{\psi_{\sigma}(x),\psi_{\sigma'}(y)\} = \delta(x-y) 
\delta_{\sigma,\sigma'}$. The basic
operators in bosonization are the charge and spin densities ($k>0$):
\begin{eqnarray}
\rho(k) &=& \sum_{p,\sigma} c^{\dag}_{p+k,\sigma} c_{p,\sigma}
\nonumber
\\
\sigma(k) &=& \sum_{p,\sigma} \sigma c^{\dag}_{p+k,\sigma} c_{p,\sigma}
\label{densities}
\end{eqnarray}
which are written as bosonic operators $b_k$ and $a_k$,
\begin{eqnarray}
b_k &=& \sqrt{\frac{\pi}{k L}} \rho(-k)
\nonumber
\\
a_k &=&  \sqrt{\frac{\pi}{k L}}\sigma(-k)
\end{eqnarray}
and obey canonical commutation relations
$[a_k,a^{\dag}_p]=[b_k,b^{\dag}_p] = \delta_{p,k}$.

In terms of the boson operators, the Kondo Hamiltonian ({\ref{almosthere}) for a single impurity becomes
\begin{eqnarray}
H &=& v_F \sum_{p>0} p (a^{\dag}_p a_p + b^{\dag}_p b_p) 
+J_z^R S^z \sum_{k>0} \sqrt{\frac{k}{\pi L}} 
(a_k + a^{\dag}_k) 
\nonumber
\\
&+& \frac{J_{\perp}}{4 \pi a} \left(S^+  
e^{\sum_{k>0} \sqrt{\frac{4 \pi}{k L}}
(a_k - a^{\dag}_k)} + h.c.\right) \, . 
\end{eqnarray}
Moreover, 
this Hamiltonian can be brought to a simpler form if one performs
a unitary transformation
\begin{eqnarray}
U = \exp\left\{S^z \sum_{k>0} \sqrt{\frac{\pi}{k L}}(a_k - a^{\dag}_k)\right\}
\label{unitary}
\end{eqnarray}
which transforms the Hamiltonian to (we drop the $b_p$ modes since
they decouple from the impurity)
\begin{eqnarray}
H' &=& U^{-1} H U = v_F \sum_{p>0} p a^{\dag}_p a_p 
+ S^z \left(J_z^R- \pi v_F \right) \sum_{k>0} \sqrt{\frac{k}{\pi L}} 
(a_k + a^{\dag}_k)
\nonumber
\\
&+& \frac{J_{\perp}}{4 \pi a} \left(S^+ + S^- \right) 
\, .
\label{2leve}
\end{eqnarray}
An important observation here
is that the unitary transformation {\it does not} affect $S^z$.
Observe that (\ref{2leve})
describes the physics of a two level system coupled to a bosonic
environment \cite{twolevel}. The basic physics of the Kondo problem
becomes rather simple from the point of view of (\ref{2leve}):
while the XY component of the Kondo interaction (associated with
the coupling $J_{\perp}$) flips the local spin and acts as a
{\it transverse field}, the Ising coupling (associated with $J_z$)
leads to a dissipative effect such that each time the spin flips
it produces particle-hole excitations at the Fermi surface. 
It can be shown that the Kondo temperature of the anisotropic
Kondo effect can be written as \cite{twolevel,costi}:
\begin{eqnarray}
k_B T_K = \frac{E_c}{\alpha} \left(\frac{\Delta_0}{E_c}\right)^{\frac{1}{1-\alpha}}
\label{tk}
\end{eqnarray}
where $E_c$ is a cut-off energy scale of the order of the bandwidth
and
\begin{eqnarray}
\Delta_0 &=& \frac{J_{\perp}}{2 \pi a}
\nonumber
\\
\alpha &=& \left(1-\frac{J^z_R}{\pi v_F}\right)^2 =
\left(1-\frac{2}{\pi} \arctan\left(\frac{\pi N(0) J_z}{4}\right)\right)^2  \, .
\label{da}
\end{eqnarray} 
Observe that
for $J_z,J_{\perp} \ll E_c$ the Kondo temperature (\ref{tk}) looks very similar to
the SU(2) expression $k_B T_K \approx E_c \exp\{-1/(N(0) J)\}$.
Indeed, from (\ref{tk}) we have
\begin{eqnarray}
k_B T_K \approx E_c \exp\left\{-\frac{\ln(1/(N(0)J_{\perp}))}{N(0) J_z}\right\} \, .
\label{tktls}
\end{eqnarray}
Notice that the Kondo temperature of an anisotropic
Kondo problem is not a single parameter quantity since it depends
on the Ising component $J_z$ and the XY component given by $J_{\perp}$.
Moreover, we have $\alpha<1 (J_z>0)$ in the case of the antiferromagnetic coupling
and $\alpha>1 (J_z<0)$ for the ferromagnetic coupling. As is well-known
the ferromagnetic Kondo effect is related with the formation of a
triplet state and therefore to the freezing of the moment (and not quenching!).

When $J^z \gg 2 v_F$ (the limit of large uniaxial anisotropy) we
see from (\ref{jzr}) that $J^z_R \to \pi v_F$ and the Hamiltonian 
reduces to 
\begin{eqnarray}
H = \frac{J_{\perp}}{2 \pi a} S_x + 
v_F \sum_{p>0} p a^{\dag}_p a_p 
\label{hextreme}
\end{eqnarray}
with the decoupling of the spin degrees of freedom to the bosonic modes.
This is the dissipationless limit of the problem.
Observe that in this limit the eigenstates of the system are
eigenstates of $S_x$, that is, the transverse field. We can immediately
see from (\ref{tk}) that in this limit
\begin{eqnarray}
k_B T_K \approx \left(\frac{\pi^2 N(0) J_z}{8}\right)^2  \frac{J_{\perp}}{2 \pi a}
\end{eqnarray}
which is a large Kondo temperature. 

In ref.\cite{grifeu} we proposed that the Kondo effect which happens
in U and Ce alloys has the structure of (\ref{hextreme}) since 
most of these systems are not cubic and therefore can be highly anisotropic.
Moreover, even in cubic systems the alloying can produce deformations
in the unit cell which can produce large local anisotropies. Thus,
if we disregard the residual coupling between the conduction electrons
and impurities the magnetic problem reduces to the transverse field
Ising model:
\begin{eqnarray}
H = \sum_{n,m} \Gamma_z(n,m) S_z(n) S_z(m) + \sum_n \Delta_R(n) S_x(n)
\label{trfisin}
\end{eqnarray}
where $\Delta_R \propto k_B T_K$ is the renormalized tunneling splitting
of the transverse field. It is obvious that the situation 
reproduces Doniach's argument: while the RKKY coupling $\Gamma_z$
works in the direction of making the local spin an eigenstate of
$S_z$ (and therefore to order it, $|\langle S_z \rangle| = 1$) the Kondo effect through $\Delta_R$ pushes the local moment
to be an eigenstate of $S_x$ and therefore leads $\langle S_z \rangle =0$.
In this picture the results for the case of insulating magnets follow
immediately and one expects power law divergences of the physical quantities
in the paramagnetic phase.

In one dimension the problem of the Kondo lattice has been studied with
the use of bosonization and has been solved exactly at a particularly
anisotropic point called the Toulouse point \cite{zek} and at half-filling
\cite{tsv}. Moreover, this problem was studied in great detail numerically
\cite{kl1dnum}. Honner and Gul\'acsi have argued that the Kondo chain
indeed maps into the transverse field Ising model, and have shown
that in the disordered case
that Griffiths singularities appear close to the transition line 
from ferromagnetic to paramagnetic behavior \cite{miklos}. This trend seems
to be reproduced in other calculations for the same problem \cite{kl1d}.
Indeed, power law behavior of the susceptibility was obtained for
the Anderson model in one-dimension with exponents very close to
the ones obtained experimentally \cite{nick}. Since the problem
of Griffiths-McCoy singularities is essentially the problem of
clusters (zero dimensional objects) surrounded by a metallic environment
it seems rather natural that (\ref{trfisin}) reproduces the
magnetic behavior of the Kondo lattice. 

The question that arises in the context of (\ref{trfisin}) is:
what is the effect of the residual interaction of the cluster with
the conduction electrons? In the paramagnetic phase we assume that
the clusters do not interact with each other. In this case one
can focus entirely on the behavior of a single cluster and its
metallic environment. This problem is actually very close to the
problem of macroscopic quantum tunneling of magnetic grains \cite{mqt}
and it is known that dissipation is a relevant perturbation to
this problem especially at low temperatures \cite{prokofev}. Consider
for instance the problem of $N$ spins in a cluster. Since we assume
the cluster to be in the ordered phase there must be two states of
the cluster which are nearly degenerate. For instance, a ferromagnetic
state with all the spins up has the same energy of a ferromagnetic
state with all the spins down (since the environment is paramagnetic
it does not bias any specific configuration). At very low temperatures 
the only way for the system
to relax is to flip all the $N$ spins at once. As we have seen
in the case of the single impurity Kondo problem (but we can
prove it to be true for the two impurity Kondo problem as well \cite{next})
requires that the XY component of the Kondo Hamiltonian to
act $N$ times over the ground state wavefunction. Since each spin flip 
requires an energy of order $J_{\perp}$ the total energy in this
case is of order $\Gamma_z (J_{\perp}/\Gamma_z)^N$ that is, the splitting
between the low lying states  of the cluster are split by
\begin{eqnarray}
\Delta_0(N) \approx \Gamma_z \exp\left\{-N \ln(\Gamma_z/J_{\perp})\right\}
\label{do}
\end{eqnarray}
and therefore is exponentially small --- as expected for the insulating
case, as well. Each time the cluster flips we expect that particle
hole excitations to be created at the Fermi surface. Since the
cluster is coupled to the electronic bath by its order parameter (magnetization
in the case of a ferromagnetic cluster or staggered magnetization in
the case of the antiferromagnetic cluster) we expect that coupling
to the bath to be {\it extensive} with the cluster size, that is,
proportional to $N$ (notice that in (\ref{2leve}) with $\alpha$ defined
in (\ref{da}) the coupling to the bath is proportional to $\sqrt{\alpha}$). 
In this case we see that the dissipation
parameter $\alpha$ has to scale like $N^2 (\tilde{J}_z/E_c)^2$ where
$\tilde{J}_z$ is the Ising coupling of the cluster to the bath
which is a function of the microscopic couplings and has to
be calculated from cluster to cluster \cite{next}. Thus, like in the case 
of a single impurity Kondo problem we can define a {\it cluster Kondo
problem} with a characteristic Kondo temperature $T_K(N)$ or tunneling
splitting $\Delta_R$ given by (using (\ref{tk}) and (\ref{do}))
\begin{eqnarray}
\Delta_R(N) = E_c \exp\left\{- \frac{\gamma N + \ln(E_c/\Gamma_z)}{1-(N/N_c)^2}\right\}
\label{tkn}
\end{eqnarray} 
where
$\gamma = \ln(J_{\perp}/\Gamma_z)$ and $N_c= E_c/\tilde{J}_z$ 
depends on the coupling constants of the problem.  
The importance of $N_c$ rests of the fact that when $\alpha>1$
there is no real Kondo effect. Thus, for $N>N_c$ the cluster
freezes and quantum fluctuations are completely suppressed. 
Indeed, for $N=N_c$ the Kondo temperature in (\ref{tkn}) vanishes.
Therefore,
$N_c$ gives the size of the largest cluster for which the
Kondo effect still takes place. We can invert (\ref{tkn}) in
order to give $N$ as a function of the splitting as
\begin{eqnarray}
\frac{N}{N_c} = \frac{1}{2 \ln(E_c/\Delta_R)}
\left(\sqrt{(\gamma N_c)^2+ 4\ln(E_c/\Delta_R) \ln(\Gamma_z/\Delta_R)  }
- \gamma N_c\right) \, .
\label{nnc}
\end{eqnarray}
Notice that there are two well defined limits of this expression
depending whether $\Delta_R$ is larger or smaller than $\Delta^*$ 
where
\begin{eqnarray}
\Delta^* = \sqrt{E_c \Gamma_z} \, \, \exp\left\{-\frac{1}{2}
\sqrt{\ln^2(E_c/\Gamma_z) + (\gamma N_c)^2}\right\} \, .
\label{dstar}
\end{eqnarray}
If $\Delta_R \gg \Delta^*$ we have $N \approx \ln(\Gamma_z/\Delta_R)/\gamma$ 
and therefore the
splitting is completely determined by $\gamma$ and we have the
same situation as in an insulating magnet. When $\Delta_R \ll \Delta^*$
and $N/N_c \approx 1$ the cluster becomes decoherent and the situation
is not described by the power law behavior. Thus, $\Delta^*$ defines
an energy scale above which power law singularities should be found
and below which a new behavior dominated by dissipation is present.
The consequences of this dissipative regime will be discussed elsewhere \cite{next}.
But a point we have to make is that since $\Delta^*$ is exponentially
dependent on $N_c$ the dissipative regime is going to be exponentially
small. Above a temperature scale $T^* = \Delta^*/k_B$  
we expect the temperature dependence of the physical quantities 
to be dominated by power law behavior. We also believe
this kind of behavior is responsible for the deviations from power
law at $T<T^*$ which is observed in some U and Ce intermetallics.

\section{Conclusions}
\label{conclusions}

In this paper we have reviewed the theoretical and experimental 
situation on the NFL behavior in U and Ce intermetallics. We argue
that the situation is inconclusive and important issues regarding
the description of the problem 
in terms either of single ion or correlated behavior
have not been solved. We argue that the Griffiths-McCoy scenario
is, so far, the only one which explains the existence of NFL behavior close
to the QCP but not exactly {\it at} the QCP. Disorder in these
systems is very important and help to pin the pieces of the ordered
phase inside of the paramagnetic phase (especially because NFL behavior
is only observed in alloys). 

We described the problem of Griffiths-McCoy singularities in
insulating magnets. This problem is now on a very firm basis,
and we know that these singularities lead to power law behavior
of the response and thermodynamic functions. We believe that
power law can describe quite well the NFL behavior observed in
a rather large temperature range in many of the systems discussed
here and especially in UCu$_{5-x}$Pd$_x$. 

We have shown that the Kondo lattice Hamiltonian can be studied
in a renormalization group sense by tracing out the higher energy
degrees of freedom deep inside or very far away from the Fermi
surface and that the effective Hamiltonian contains the 
basic ingredients required for the {\it local} Doniach description
of these systems. We argued on the basis of the mapping of
the single impurity Kondo problem into the dissipative two level
system that in the limit of high
local magnetic anisotropy the Kondo problem indeed maps into
a transverse field Ising model, which has been shown to
present Griffiths-McCoy singularities in its phase diagram.
We also have argued that the power law behavior disappears
at temperatures smaller than $T^*=\Delta^*/k_B$
(which is probably quite small) where the situation is dominated
by dissipative physics. What we have shown, therefore, is that
like in the Kondo disorder picture \cite{ucupdnmr,miranda} there
is a distribution of Kondo temperatures which is {\it not} of
single ion character but has to do with the Kondo temperature
of isolated clusters. Observe that the distribution is {\it not} arbitrary
but determined completely by the statistical distribution of
clusters in a percolation problem. If a residual interaction
between clusters exists close to the QCP then with the lowering
of the temperature a
quantum spin glass state is a possible ground state \cite{huse}. 
In this
case a return to Fermi liquid behavior with a strong temperature
crossover is expected.
Otherwise,
if the clusters are truly non-interacting then the quantum
super-paramagnetic state dominated by Griffiths-McCoy singularities
can exist and real singularities in the response functions
must be observed.

I would like to thank my collaborators, B.~A.~Jones, G.~Castilla, 
M.~de Andrade, B.~Maple and D.~MacLaughlin
for many illuminating discussions. I also would like to acknowledge D.~Cox, 
P.~Coleman, V.~Dobrosavljevi\'c, G.~Kotliar,
A.~Millis, E.~Miranda and A.~Tsvelik for their constructive criticism. 
Many illuminating discussions with I.~Affleck, 
M.~Aronson, V.~Barzykin, W.~Beyermann,   
B.~Coqblin, M.~Gul\'acsi, L.~De Long, N.~Kioussis, J.~Mydosh, 
N.~Prokof'ev, H.~R.~Ott, T.~Senthil and
G.~Stewart on many aspects of the NFL are greatly
appreciated. I thank H.~Rieger for many discussions
on the problem Griffiths-McCoy singularities. 
I also would like to thank E.~Bascones, F.~Guinea 
and the organizers of the VI International
Summer School ``Nicol\'as Cabrera'': R.~Villar, 
F.~G.~Aliev, J.~C.~G\'omez-Sal and
H.~Suderow, for their hospitality during the Summer School.
I acknowledge partial support from the Alfred P.~Sloan foundation
and a Los Alamos CULAR grant under the auspices of the U.~S.
Department of Energy.

\begin{figure}
\epsfxsize=6cm
\epsfysize=6cm
\hspace{1cm}
\epsfbox{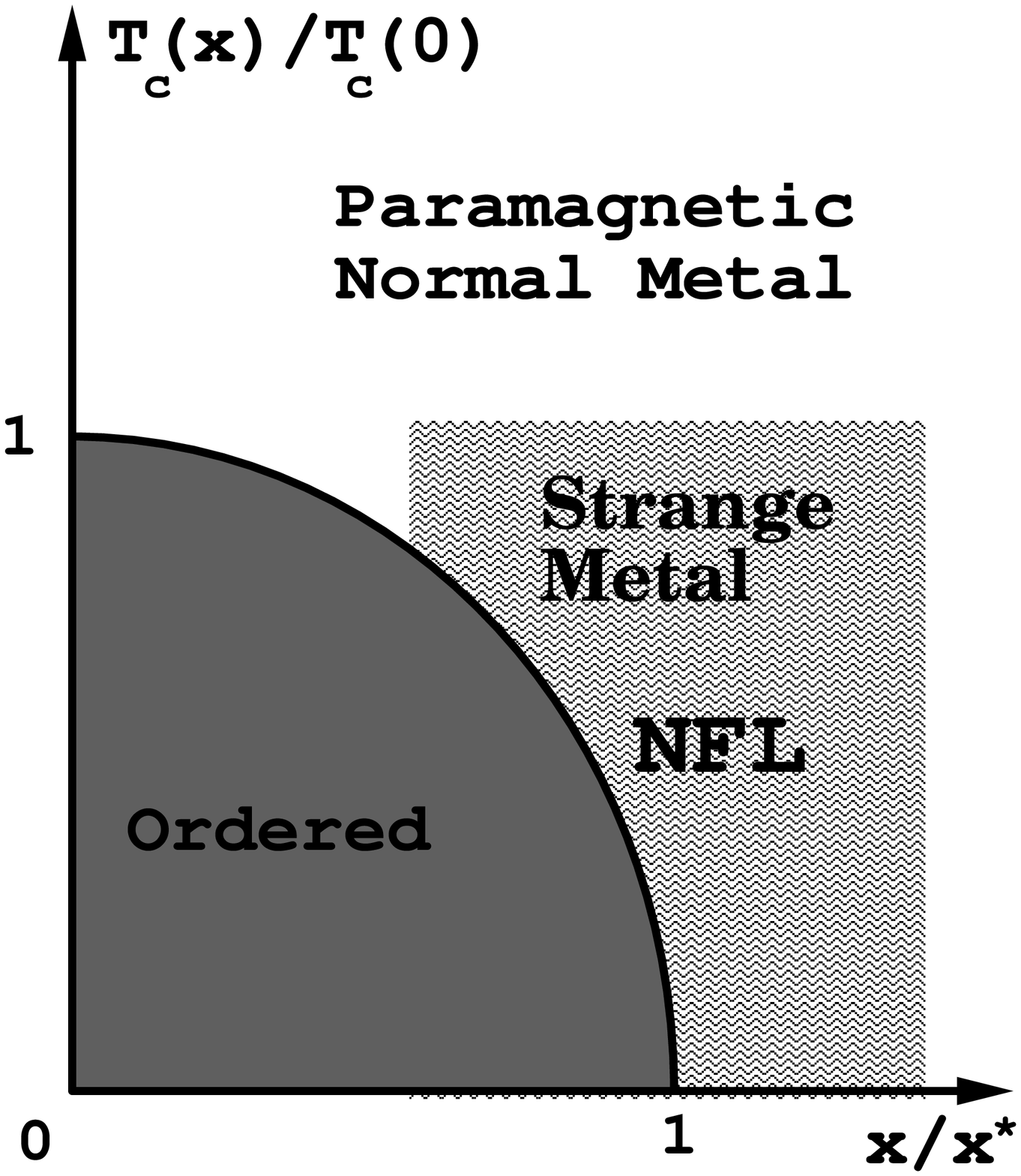}
\caption{Typical phase diagram for heavy fermion alloys. Vertical axis
is the critical temperature of the ordered phase and the horizontal axis
is chemical composition.}
\label{phdi}
\end{figure}

\begin{figure}
\epsfxsize=6cm
\epsfysize=6cm
\hspace{1cm}
\epsfbox{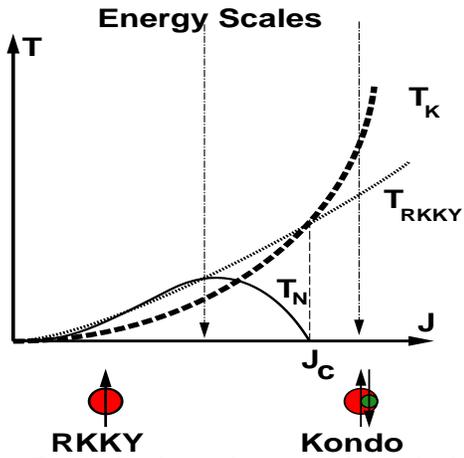}
\caption{Doniach phase diagram: long dashed like is the Kondo temperature, $T_K$;
short dashed line is the RKKY temperature,$T_{RKKY}$; the continuous line is the
ordering temperature $T_N$.}
\label{doni}
\end{figure}

\begin{figure}
\epsfysize10cm
\hspace{1cm}
\epsfbox{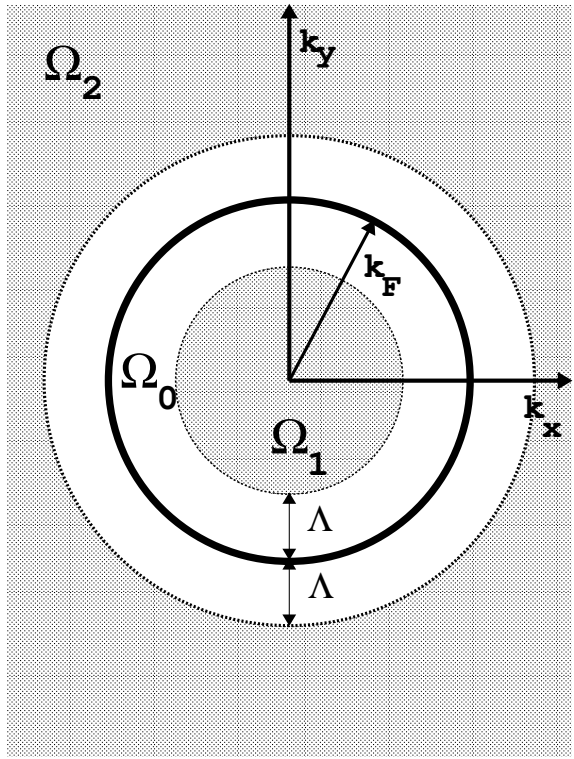}
\caption{Different regions of energy close to the Fermi surface.}
\label{fs}
\end{figure}

\end{document}